\documentclass[a4paper,12pt]{article}

\usepackage{graphicx} 
\usepackage{geometry}
\usepackage{setspace}
\usepackage{verbatim}
\usepackage{booktabs}
\usepackage{amsmath}
\usepackage{bbold}
\usepackage{authblk}
\usepackage{threeparttable}
\usepackage{tabularx}
\newcolumntype{Y}{>{\raggedright\arraybackslash}X}
\usepackage[maxbibnames=99, style=numeric-comp, sorting=none]{biblatex} 

\usepackage[hidelinks]{hyperref}
\usepackage[nameinlink,capitalise]{cleveref}

\title{Network-based modeling of cocaine trafficking flows and displacement effects}

\author[1]{H.A. (Bart) Peters\thanks{H.A.Peters@tudelft.nl}}
\author[1]{Frederike Oetker}
\author[2]{Robby Roks}
\author[3,4]{Roy Lindelauf}
\author[1]{Robbert Fokkink}
\author[1]{Huijuan Wang\thanks{H.Wang@tudelft.nl}}

\affil[1]{Faculty of Electrical Engineering, Mathematics and Computer Science, Delft University of Technology, Delft, The Netherlands}
\affil[2]{Erasmus School of Law, Erasmus University Rotterdam, Rotterdam, The Netherlands}
\affil[3]{Faculty of Military Sciences, Data Science in Military Operations, Netherlands Defence Academy, Breda, The Netherlands}
\affil[4]{Tilburg School of Humanities and Digital Sciences, Tilburg University, Tilburg, The Netherlands}

\date{\today}

\begin{document}

\maketitle

\section{Abstract}
The worldwide cocaine market is undergoing an extraordinary surge. Insights on how cocaine is trafficked from production countries to consumer markets are limited, and often based on observed trafficking routes (e.g., seizures) alone. In this paper, we introduce a network-based model of actual transnational cocaine flows, beyond flows seized. In this model, cocaine is routed through an international transportation network of land and sea connections, where each link is assigned an interception risk metric for containerized transport. This interception risk metric combines features related to law enforcement inspections, containerized trade, and transport connectivity into three key driving components of interception risk using Principal Component Analysis (PCA).
The interception risk metric on each link is then constructed from these driving components, where the importance of each component is characterized by a strength parameter. Given a set of strength parameters, cocaine is assumed to be routed from production countries to consumption countries via routes that minimize the total risk of interception, subject to country-specific supply and demand constraints.
In the end, the strength parameters are calibrated by comparing the corresponding flow outcomes with known trafficking activity.
From these final trafficking flows associated with the calibrated parameters, we are able to identify `blind spots' (e.g., nonzero flows on links with small or none observed trafficking activity) and provide more evidence of the waterbed effect (e.g., changing routes as a result of increased interdiction efforts by law enforcement).

\section{Introduction}
In recent years, the global cocaine market has experienced unprecedented growth. Annual production volumes in South America continue to increase, with no indication that maximum production capacity has yet been reached \cite{UNODC2023Cocaine}. At the same time, an increase in cocaine demand is observed in Europe, the world's largest consumer market together with the United States \cite{UNODC2025WorldDrugReport, EUDACocaineAnalysis2022}.
After being produced in the Andean region, with the main production countries being Colombia, Peru, and Bolivia, cocaine is mainly trafficked to consumer markets by container ships originating from ports in Central and South America \cite{Baarsma2022DenkwerkKrijgen, UNODC2021The1}. However, cocaine traffickers have also recently renewed interest in West African ports as transit hubs to circumvent customs inspections in destination ports in Europe \cite{INTERPOL2023CocaineAfrica, VanDenEeden2026GeorganiseerdeCriminaliteit}.
\\
\\
Smuggling methods further illustrate the sophistication and adaptability of cocaine trafficking groups, a pattern also reflected in the strong growth of global seizure volumes and the central role of European maritime gateways. Global cocaine seizures increased from roughly $630$ tonnes in 2010 to over $2200$ tonnes in 2023, with Western and Central Europe rising from approximately $61$ tonnes to more than $440$ tonnes during the same period, indicating the growing importance of European consumer markets and transit hubs \cite{UNODC2025WorldDrugReport}. Large container ports play a particularly prominent role: seizure quantities linked to hubs such as Rotterdam, Antwerp, Hamburg, and major Spanish ports, such as Algeciras and Valencia, show persistent interception activity, with Rotterdam alone registering seizure figures exceeding $70$ tonnes in 2021, while Antwerp and Iberian ports account for similarly substantial volumes \cite{Baarsma2022DenkwerkKrijgen, EUDA2025EuropeanDrugReport}. Furthermore, between $80\%$ and $90\%$ of cocaine seizure volumes are directly linked to maritime trafficking in recent years \cite{UNODC2025WorldDrugReport}.
\\
\\
Given the sheer volume of containers handled by ports, only a small fraction can be inspected by authorities on a daily basis \cite{Europol_CriminalNetworks_EUPorts_2023}. In turn, this creates a probabilistic setting for cocaine traffickers where they can embed illicit packages within licit trade flows \cite{Sergi2022Pacman}.
Various trafficking techniques have been used, including concealment within commercial container shipments, infiltration of logistics chains through corrupt insiders, maritime transfers between vessels at sea, and the use of encrypted communication to coordinate distribution \cite{Europol_CriminalNetworks_EUPorts_2023}. 
Smaller-scale methods such as body packing, hidden compartments in vehicles, and parcel shipments complement large-volume maritime trafficking, enabling diversification of risk and maintaining supply continuity even when individual routes are disrupted. 
\\
\\
Criminal infiltration and its resulting governance challenges have already been widely studied for individual ports in Belgium \cite{Easton2020NodalAntwerp}, Italy \cite{Antonelli2021ExplorationActivities, Antonelli2024UnpackingPorts}, Greece \cite{Sergi2024CocainePiraeus}, The Netherlands \cite{Bisschop2019UitdagingenHaven, Staring2019DrugscriminaliteitFenomeen, Roks2021GettingRotterdam, Staring2023DrugApproach} and the southern cone of South America \cite{Sampo2022CocaineUruguay, Sampo2025RoleEurope}. Collectively, these studies show that container ports are not merely passive transit points, but structurally embedded areas where licit and illicit trade flows intersect \cite{VandeBunt2014EmbeddednessOC}.
Ultimately, the combination of rising seizure volumes, concentration around high-capacity container ports, and diversification of concealment strategies highlights how cocaine trafficking maintains both logistical efficiency and operational resilience, reinforcing its persistent challenge for public health systems and law enforcement agencies.
\\
\\
Although annual estimates exist for the amount of cocaine produced and consumed globally \cite{UNODC2023Cocaine}, actual cocaine trafficking flows remain poorly understood. Currently, only partial information on these illicit flows is obtained from law enforcement intelligence, such as the amount of cocaine seized per container port or transportation route. 
This information has already been used to map and study total cocaine trafficking flows.
For example, in the most widely-used method for reconstructing cocaine flows between countries, first introduced by the United Nations Office on Drugs and Crime (UNODC) \cite{UNODC2015DrugRoute} and later proposed by Aziani \cite{Aziani2018IllicitEstimation}, incoming cocaine trafficking flows to each country are thought to be proportional to the quantities intercepted on each incoming link, provided that the country has reported their individual drug seizure (IDS) statistics to the UNODC \cite{Giommoni2022_InterdictingInterventions, Aziani2021QuantitativeEurope, Berlusconi2017DeterminantsApproach, Giommoni2017HowEurope}. However, countries report their drug seizure statistics voluntarily to the UN, meaning that drug flows constructed from this method could be biased towards reporting countries \cite{Underwood2023SeizureData}. 
In addition, due to the heterogeneity of drug inspections from law enforcement across countries, large drug flows might be directed to countries with strong law enforcement capabilities using this method, thereby not capturing possible drug flows to countries with poor law enforcement.
\\
\\
Efforts have been made to improve this drug trafficking flow construction method by supplementing the IDS database with link-level seizure data from secondary sources, such as governmental intelligence reports \cite{Screen2025TransnationalConstruction, Boivin2013DrugEconomy, Boivin2014RisksEconomy, Boivin2011MondeIllicites}. However, in the methods mentioned above, drugs are only allowed to flow on links where seizures have occurred or intelligence is gathered, thereby not accounting for possible undetected trafficking routes (e.g., blind spots). 
In addition, insights collected from seizure and intelligence data are heavily dependent on resources and priorities of law enforcement agencies, which can vary substantially between countries \cite{EUDACocaineAnalysis2022, UNODC2021The1, Sergi2022Pacman}, as well as the reporting rate of the respective country \cite{Underwood2023SeizureData}.
\\
\\
Several models have been suggested to account for potential drug transports on routes without any observed trafficking activity in the construction of drug flows.
For example, the directionality of cocaine trafficking flows between countries has been modeled based on an increase in the wholesale price of cocaine \cite{Chandra2011InferringData, Chandra2013WhatEurope, Chandra2015TransnationalComparison}. In these price-driven models, cocaine flows exist between countries whose prices are substantially correlated. However, these models are constrained by a weak representation of network logistics and have previously been employed in a limited geographical scope only.
Other agent-based models for reconstructing cocaine trafficking flows are based on risk-profit mechanisms of licit supply chains in Central America \cite{Magliocca2024TowardsModeling, Magliocca2019ModelingSystem}. Here, trafficking flows are constructed as a trade-off between profit maximization and risk management, where risk is largely determined from seizure information on trafficking routes. 
Lastly, global cocaine trafficking flows have been constructed on trajectories of container vessels using a network flow optimization approach \cite{Leibbrandt2023_DrugNetwork}. Here, country-level trafficking flow estimates were obtained by minimizing the number of transport connections used to traffic cocaine from production to consumption countries. In reality, however, cocaine traffickers might exploit indirect routes to decrease the risk of cocaine interception rather than relying on direct transportation routes \cite{Magliocca2021ComparativeSand}.
Hence, existing models for estimating drug trafficking flows are often limited in geographical scope, tend to substantially rely on seizure information, fail to incorporate transport logistics, and/or struggle to jointly account for flow changes resulting from interdiction efforts.
\\
\\
Changes in maritime cocaine seizures over time often lead to claims of the so-called `waterbed' effect (also known as the balloon effect \cite{Windle2012PoppingHypothesis, Magliocca2019ModelingSystem}). This effect refers to the spatial displacement of trafficking routes in response to intensified law enforcement efforts \cite{Bagley2013EvolutionDrugTrafficking, reuter_mobility_drug_trafficking_2014}. However, since seizures only show the observed part of cocaine trafficking flows, additional information on these flows is needed to support these claims of the waterbed effect.
\\
\\
To address these gaps in the construction of cocaine trafficking flows and the claims of the waterbed effect, we propose a network flow optimization approach that models cocaine trafficking on an integrated transportation network (combining maritime shipping and road connections) while minimizing the risk of cocaine interception. 
For this approach, we conceptualize transnational cocaine trafficking as an economic process in which drugs move from production countries to consumption countries through a chain of intermediaries, assuming that cocaine trafficking costs are primarily driven by compensation for interception risk \cite{Boivin2013DrugEconomy}. 
Furthermore, the waterbed effect can be studied by manually varying law enforcement inspections, directly resulting in increased risk of interception on certain links.
\\
\\
We will first construct the legal transportation network, used by criminal organizations to traffic cocaine. This transportation network is a network of countries that are connected by a link if they share a land border or container shipping connection. Hence, we focus on cocaine trafficking on land and sea routes, as the majority of cocaine is expected to be transported this way \cite{UNODC2025WorldDrugReport}. Secondly, we will estimate the amount of cocaine produced, consumed, and seized for each county, based on data from the UNODC's annual World Drug Reports (WDRs) \cite{UNODC2025WorldDrugReport}, using a combined supply and demand approach \cite{Giommoni2022_InterdictingInterventions}.
Thirdly, we will construct a seizure-independent interception risk metric for each link as a function of several features related to transport connectivity, law enforcement inspection strategies, corruption vulnerability, and containerized trade.
In our model, cocaine will then flow from production countries to consumption countries using the optimal route that minimizes the overall risk of cocaine interception on the network, whilst adhering to constraints on the supply and demand for each country. 
\\
\\
By analyzing how the risk of cocaine interception is perceived by criminals, we will be able to study how cocaine is most likely to flow around the world. In turn, this will allow us to provide insights into possible blind spots in the cocaine supply chain by comparing the resulting flow with known seizure information. Seizures on trafficking routes will therefore only be used for external validation in this work. In addition, we will provide an explanation of the waterbed effect by studying changes in cocaine trafficking flows resulting from an increase in law enforcement inspections on selected trafficking routes.
\\
\\
This paper is structured as follows. In Section \ref{sec_network} we will describe the construction of the transportation network used for trafficking cocaine.
After that, we will propose the cocaine trafficking model in Section \ref{sec_routing}. Here, we will present an overview of the relevant features used to construct the interception risk metric, as well as the trajectory routing mechanism, based on this metric. Model performance and results on blind spot detection and flow displacements will then be discussed in detail in Section \ref{sec_results}, followed by a conclusion in Section \ref{sec_conclusion}.

\section{Transportation network construction}
\label{sec_network}
\begin{figure}[h]
\centering
\includegraphics[width=1\textwidth]{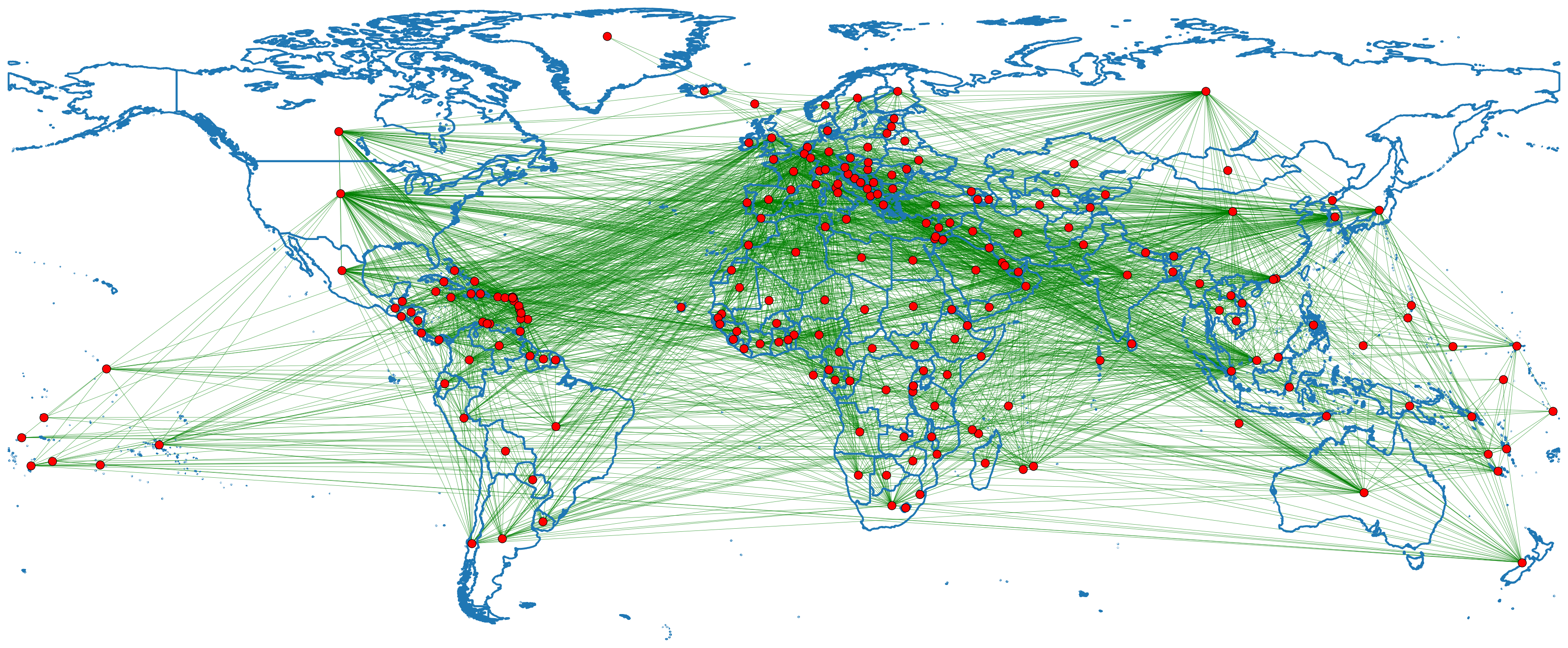}
\caption{The transportation network displayed on the world map.}
\label{fig_ctn}
\end{figure}

Our objective is to model global cocaine trafficking using a network flow model. In this model, cocaine is trafficked via a legal transportation network that contains land and sea connections between countries. Cocaine trafficking on airplanes will be neglected, as air routes only account for a small fraction of reported volumes, while maritime and land routes carry bulk consignments \cite{INTERPOL2023CocaineAfrica, UNODC2025WorldDrugReport}. 
In the transportation network,  two countries (nodes) are connected by a link (edge) if they share a land border or there exists a maritime container shipping connection between them. To determine the latter, we will consider all shipping connections between countries that are present in the Global Liner Shipping Network (GLSN) at the country level \cite{KojakuMultiscaleNetwork2019, XuEstimatingNetworks2020, Xu2020ModularScience}. 
\\
\\
The country-level GLSN is constructed based on all $1622$ container liner shipping service routes of the world's top $100$ liner shipping companies in 2015 \cite{XuTransportationNetworkData2022}. These service routes contain information on the trajectory of container vessels and are relatively stable over time \cite{Leibbrandt2023_DrugNetwork}. The trajectory information of a service route describes the sequence of port calls from origin port to destination port, including all ports in between. The service routes are bidirectional, meaning that container vessels travel back and forth between the first and last port on the route. Because of this, each service route can be mapped to a complete subgraph, such that any two ports in the service route are connected by a link \cite{Xu2020ModularScience}.
This procedure then results in a connected port-to-port network consisting of $977$ unique ports and $16680$ inter-port connections. In the country-level GLSN, two countries are connected by a sea-link if at least two ports from both countries are connected in the port-to-port network. The countries used in this work have been named based on the list of standard country and area codes from the United Nations Statistics Division (UNSD) \cite{UNSD2025StandardM49}. The country-level GLSN can then be considered as the sea-link layer of our transportation network.
\\
\\
To construct a network layer of country-to-country road connections, we will consider the GeoDataSource Country Border dataset \cite{GeoDataSource2025CountryBordersDatabase}. In this layer, two countries are connected by a land-link if they share a land border. The same UNSD definition of countries is used.
We will then construct the transportation network by considering the union of the sea-link and land-link layers.
This process results in an undirected transportation network $G=(V,E)$, where $V$ and $E$ are the sets of nodes and links, respectively, which has $N=|V|=224$ nodes and $|E|=11268$ links in total. The transportation network $G$, as displayed in Figure \ref{fig_ctn}, can then be represented by its adjacency matrix $A$ with entries $a_{ij}$, which are equal to one if there exists a link from country $i$ to $j$ and equal to zero otherwise. 
Each existing network link has retained a label that specifies if the link is present in the sea layer, road layer, or both.

\section{Cocaine trafficking model}
\label{sec_routing}
Our model of cocaine trafficking assumes that cocaine is routed from production countries to consumption countries via the transportation network, minimizing the risk of interception by law enforcement whilst adhering to constraints on the supply and demand for each country.
This minimization of risk should therefore be considered as a simplification of the trafficking decision process. The supply and demand for each country in each year are derived from UNODC data, following the approach given in Section \ref{sec_appendix_marketparametercalculation}.

\subsection{Interception risk of a link}
We define interception risk for a unit traffic when traversing a directed link ($i$,$j$) in the transportation network in a year $t$ to be proportional to the probability of being inspected or sampled by law enforcement and customs authorities when arriving in country $j$ in that year. This interception risk is the risk perceived by criminals rationally and may depend on several features of the link ($i$,$j$) and destination country $j$. We will first introduce the relevant features in Section \ref{sec_interception_risk_determinants} and then model the interception risk using these features in Sections \ref{sec_PCA} and \ref{sec_interception_risk_modeling}.

\subsubsection{Determinants of interception risk}
\label{sec_interception_risk_determinants}
The interception risk of link ($i$,$j$) in year t may depend on the following features of the link and of country $j$ in the same year. These features are provided in Table \ref{tab:variablesRisk}, and are defined and motivated as follows. 
\begin{table}[h]
\centering
\begin{threeparttable}
\caption{Features used to calculate each link's interception risk.}
\label{tab:variablesRisk}
\begin{tabularx}{\linewidth}{
    >{\hsize=0.5\hsize\raggedright\arraybackslash}X
    >{\hsize=1.5\hsize\raggedright\arraybackslash}X
    c} 
\hline
\textbf{Feature} & \textbf{Feature description on each link ($i$,$j$)} & \textbf{Time period} \\
\hline
Container ship movements$^{*}$ & Number of unique container ships traveling from $i$ to $j$. & 2015 \\
Container port throughput$^{\dagger}$ & Annual volume of containers handled by the ports in $j$. & 2010--2023 \\
Production proximity & Indicator related to the proximity of the link to the cocaine production region. & static \\
Control of Corruption$^{\ddagger}$ & Perceptions of the extent to which public power is exercised for private gain in $j$ \cite{KaufmannKraayMastruzzi2010}. & 2010--2023 \\
Overland connectivity & Boolean variable indicating if the link does not correspond to a road connection. & static \\
Intraregional connectivity$^{\S}$ & Boolean variable indicating if $i$ and $j$ are not within the same region. & static \\
Price difference$^{\S}$ & Difference in wholesale price between $j$ and $i$. & 2010--2023 \\
\hline
\end{tabularx}
\begin{tablenotes}
\footnotesize
\item[*] Source: Transportation network \cite{XuTransportationNetworkData2022}.
\item[$\dagger$] Source: UNCTAD \cite{UNCTAD2025ContainerPortThroughput}.
\item[$\ddagger$] Source: World Bank (Worldwide Governance Indicators) \cite{WorldBank2025WGI}.
\item[$\S$] Source: UNODC \cite{UNODC2025WorldDrugReport}.
\end{tablenotes}
\end{threeparttable}
\end{table}
\\
\\
The \textit{Container ship movements} of link $(i,j)$ are the number of unique container ships traveling from $i$ to $j$ and is zero if $i$ is connected to $j$ solely by land connection. Given the law enforcement inspection capacity of country $j$, a larger number of incoming container ships could lead to a reduced inspection probability \cite{Europol_CriminalNetworks_EUPorts_2023, Eski2011PortSecurity}. 
\\
\\
Similarly, the interception probability tends to be small if the \textit{container port throughput} (CPT) of country $j$ is large \cite{Europol_CriminalNetworks_EUPorts_2023}. This throughput includes the number of loaded, unloaded, and transshipment containers of a port, and is therefore a measure of the quality of port infrastructure and trade connectivity. 
Annual country-level CPT is obtained from the United Nations Trade and Development (UNCTAD) database \cite{UNCTAD2025ContainerPortThroughput}, given in terms of the total number of twenty-foot equivalent (TEU) sea containers handled by all container ports in the country.
Missing values are determined from linear interpolation and extrapolation of a country's own time series. Values for landlocked countries or small island states that did not report any numbers are imputed as zero.
\\
\\
Containers from high-risk origin countries are more likely to be selected for inspection by seaport customs officials \cite{Pourakbar2018RoleChains}. For example, containers originating from Colombia and Peru are often sampled by port authorities upon arrival, as cocaine is manufactured in both countries. Consequently, criminals often traffic their cocaine to another country close to the production region first before sending it to Europe or the United States. At the same time, port authorities are aware of this and therefore tend to conduct more frequent inspections on containers originating from South America, the Caribbean and Central America (including Mexico), resulting in a `cat-and-mouse' dynamic between criminal organizations and port officials \cite{Sampo2022CocaineUruguay, EUDACocaineAnalysis2022, Magliocca2019ModelingSystem, Sergi2022Pacman}. 
We propose a \textit{production proximity} metric $r_{ij}$ for each link ($i$,$j$) as
\begin{equation}
    \label{eq:customsoriginsamplingrisk}
    r_{ij} = 
    \begin{cases}
        r_i-r_j & \text{if }\ r_i\geq r_j, \\
        0 & \text{else}.
    \end{cases}
\end{equation}
where $r_i$ is a risk indicator based on the proximity of country $i$ to the region of cocaine production. Concretely, we assign $r_i=2$ if $i$ is a production country, $r_i=1$ if $i$ is another country in South America, the Caribbean or Central America (including Mexico), and $r_i=0$ otherwise. A high $r_{ij}$ implies a high probability of inspection by law enforcement due to the proximity of both countries to the production region. 
\\
\\
Container inspections can also be bypassed by corrupting port personnel, such as terminal operators, customs officials, or truck drivers \cite{Europol_CriminalNetworks_EUPorts_2023}. In the port of Rotterdam, for example, corruption of port workers is thought to be essential for smuggling drugs \cite{Roks2021GettingRotterdam}. In addition to that, previous studies showed that corrupted countries are generally more likely to be a transit country for drug trafficking \cite{Trumbore2014SmugglerDataset}.
Therefore, the interception probability of link ($i$,$j$) tends to be small if country $j$ is vulnerable to corruption, which is approximated by country $j$'s \textit{Control of Corruption} (CoC) measure, as recorded in the World Bank's Worldwide Governance Indicators \cite{WorldBank2025WGI}. It describes for each country the annual perceptions of the extent to which public power is exercised for private gain, including both petty and grand corruption, as well as state capture by elites and private interests \cite{KaufmannKraayMastruzzi2010}. Countries with a high CoC score have strong corruption control, resulting in a higher interception risk in our model. 
\\
\\
Land borders are often less controlled by law enforcement compared to sea borders, making it relatively easy to smuggle large batches of contraband over the border \cite{reuter_mobility_drug_trafficking_2014}. Trafficking groups are also adaptive in picking suitable border crossings. In Central America, for example, criminals tend to explore new land corridors with low population density, thus avoiding regions with high local interdiction rates \cite{MAGLIOCCA2022ShiftingInterdiction}. For this reason, an \textit{overland connectivity} feature is incorporated in our model as a boolean variable. This feature is equal to $1$ if $i$ is connected to $j$ via a road connection, leading to lower risk of interception, while it is $0$ otherwise.
\\
\\
Law enforcement control also depends on whether both countries are within the same geographical region or not. For example, a container transported from Spain to Germany is not likely to be sampled by German customs officials, since Spain and Germany are both within the same region. At the same time, border control between Mexico and the United States is relatively strong, despite the link being an overland connection.
Therefore, an boolean feature \textit{intraregional connectivity} will be considered. It is equal to $1$ ($0$) if $i$ and $j$ are (not) within the same geographical region, suggesting a low (high) interception risk. We use the regional classification proposed by the UNODC in their WDRs, with the exception that we consider the production countries (e.g., Bolivia, Colombia, and Peru) as a separate group called the Andean region \cite{UNODC2025WorldDrugReport}. 
\\
\\
Lastly, the risk of interception on a link ($i$,$j$) could be related to the wholesale \textit{price difference} between $j$ and $i$. Cocaine prices tend to be higher in countries where inspection risks, due to law enforcement activity, are higher \cite{Reuter1986RisksEnforcement,Boivin2014RisksEconomy}.
Consequently, cocaine trafficking groups might perceive higher risk of interception on network links with large price differentials.
Therefore, the difference in wholesale price between $j$ and $i$ of link ($i$, $j$) will be considered as a determinant of the link's interception risk. The difference is taken as zero if the wholesale price of cocaine in country $i$ is higher than in country $j$. Wholesale price data are obtained from the UNODC WDRs \cite{UNODC2025WorldDrugReport}, and missing values are imputed similarly to the imputation of wholesale purities as described in Section \ref{sec_appendix_marketparametercalculation}.

\subsubsection{Principal Component Analysis}
\label{sec_PCA}
The features related to interception risk might be correlated. These correlations can be used to create clusters of variables using Principal Component Analysis (PCA), thereby reducing the total number of variables used to model each link's interception risk. In PCA, features are linearly transformed into orthogonal Principal Components (PCs) that capture maximum variance \cite{Jolliffe2002PCA}. In criminological applications, it has been used to extract driving factors related to crime patterns \cite{KhanyileAdeliyiAroba2025PCA} and risks of money laundering \cite{Riccardi2025DataCorruption}.
In our case, we use data of the features displayed in Table \ref{tab:variablesRisk} for all available years as input for the PCA. If a feature only has available data for one year, the same data will be used for the other years.
In PCA, the scores (e.g., the new feature coordinates in PC space) are calculated for each year as
\begin{equation}
    \label{eq_PC}
    y_{l,ij} = \sigma_k\sum^{n}_{k=1}q_{lk}z_{k,ij},
\end{equation}
where $z_{k,ij}$ is a standardized feature on link ($i,j$), $n=7$ equals the number of unique features, and $q_{lk}$ are the time-invariant loadings, indicating the positive influence of standardized feature $z_k$ on score $y_l$. Also, to ensure that each feature contributes positively to interception risk, we consider $\sigma_k=-1$ if $z_k$ corresponds to container ship movements, CPT, overland connectivity or intraregional connectivity, while $\sigma_k=1$ for the other features, in line with their definitions presented in Table \ref{tab:variablesRisk}. In this way, positive loadings $q_{lk}$ result in high interception risk for large scores $y_{l,ij}$.
\\
\\
The loadings, together with the PCA's variance statistics, are shown in Table \ref{tab:pca_compact} for the first three PCs, which are the only components showing an eigenvalue greater than one. These three PCs cover $55.8\%$ of the total variance of all features and will later be used in the calculation of the interception risk metric.
\begin{table}[htbp]
\centering
\caption{Principal Component Analysis of interception risk determinants.}
\label{tab:pca_compact}
\begin{tabular}{lccc}
\hline
\textbf{Components} & \textbf{PC 1} & \textbf{PC 2} & \textbf{PC 3} \\
\hline
\textbf{Interpretation} & \textbf{Proximity} & \textbf{Maritime} & \textbf{Economic} \\
 & \textbf{} & \textbf{trade} & \textbf{transport} \\
\hline
\textbf{Variance} &   &   &   \\
\hline
Eigenvalue                     &  $1.65$      &    $1.23$    &   $1.02$     \\
Proportion of overall variance &   $23.6\%$     &   $17.6\%$     &   $14.6\%$     \\
Proportion of model variance   &   $42.3\%$     &   $31.5\%$     &   $26.2\%$     \\
\hline
\textbf{Loadings} &   &   &   \\
\hline
Intraregional connectivity      & 0.532 & 0.234 & -0.350 \\
Overland connectivity           & 0.473 &       & -0.485 \\
Production proximity & 0.437 &       & 0.399  \\
Price difference      & 0.389 &       & 0.567  \\
Control of corruption  & 0.336 & -0.325 & 0.231 \\
Container ship movements     &       & 0.713 &        \\
Container port throughput     &       & 0.560 & 0.325  \\
\hline
\end{tabular}
\begin{flushleft}
\footnotesize
Notes: Only loadings with absolute value greater than $0.2$ are reported. Proportion of model variance is computed relative to the retained components.
\end{flushleft}
\end{table}
\\
The three PCs can be interpreted as latent structural dimensions, serving as key drivers of interception risk. 
The first PC primarily combines proximity-related features of a link, such as intraregional and overland connectivity and production proximity.
The second PC is mainly associated with maritime trade, reflected in container ship traffic and container throughput along each link. 
The third PC reflects economic transport, as it is mainly constructed from the difference in wholesale price and production proximity.

\subsubsection{Interception risk perception}
\label{sec_interception_risk_modeling}
Through PCA, the original $n=7$ features of each link ($i,j$) are reduced to $m=3$ scores. 
Each score $y_{l,ij}$, where $l\in[1,m]$, will be linearly rescaled to a range of positive values $[a,b]$ as
\begin{equation}
    \hat{y}_{l,ij}=a+(b-a)\frac{y_{l,ij}-\min_{(i,j)}y_{l,ij}}{\max_{(i,j)}y_{l,ij}-\min_{(i,j)}y_{l,ij}},
\end{equation}
where $(a,b)=(1,10)$ will be considered as an example in this work.
The interception risk, or weight, on link ($i,j$) is then modeled as
\begin{equation}
    \label{eq_w}
    w_{ij} = \prod^{m}_{l=1}\hat{y}_{l,ij}^{\alpha_l},
\end{equation}
where $\alpha_l$ is a non-negative strength parameter that controls the influence of $\hat{y}_{l,ij}$ on $w_{ij}$.
For a large $\alpha_1$ value, for example, interception risk will be relatively small for links with low proximity scores.
\\
\\
The assumption that each PC contributes positively to interception risk is supported by the interpretation of the original $n=7$ features in Table \ref{tab:variablesRisk}, the positive rescaling in Equation (\ref{eq_PC}) and the observation that the loading vectors, as given in Table \ref{tab:pca_compact}, contain mainly positive elements. Rescaling each score $y_{l,ij}$ to $\hat{y}_{l,ij}$ within a positive range $[a\geq1,b]$ ensures that all PCs are presented on the same scale and, in turn, each strength parameter directly reflects the importance of its corresponding PC in the calculation of interception risk.
\\
\\
Our cocaine trafficking model assumes that cocaine traffickers conduct their criminal risk analysis and decide the interception risk ${w_{ij}}$ of all links by weighing those seven features. This choice can be approximately mapped to an equivalent parameter set $(\alpha_1$,$\alpha_2$,$\alpha_3)$.

\subsection{Identifying trafficking flows}
Given the interception risk $\{w_{ij}\}$ of all links in the transportation network, cocaine traffickers are assumed to determine their trafficking paths that minimize the total interception risk, while supply and demand should be matched in each country.
If a trafficking path $\mathcal{P}$ from country $i$ to $j$ on the transportation network consists of only one hop, interception cost is given by $w_{ij}\varphi\cdot f_{ij}$ where the interception probability $w_{ij}\varphi$ is assumed to be proportional to interception risk $w_{ij}$, and $f_{ij}$ represents the flow volume on link ($i$,$j$). 
If a path $\mathcal{P}$ consists of multiple hops, the probability that trafficking does not encounter any interception is 
\begin{equation}
    \prod_{(i,j)\in \mathcal{P}}1-w_{ij}\varphi.
\end{equation}
When $\varphi\to 0$, thus the interception probability is small,  $\prod_{(i,j)\in \mathcal{P}}(1-w_{ij}\varphi)\approx1-\sum_{(i,j)\in \mathcal{P}}w_{ij}\varphi$. In this case, the interception cost for a flow of volume $f_{ij}$ along path $\mathcal{P}$ is approximately $\sum_{(i,j)\in P}w_{ij}\varphi\cdot f_{ij}$. 
Our model then identifies trafficking flows $\{f_{ij}\}$ as those minimizing the total interception cost: 
\begin{equation}
    \label{eq_flowoptimizationproblem}
    \min \left\{ \sum_{(i,j)\in E}w_{ij}f_{ij} \right\},
\end{equation}
while supply and demand should be matched in each country $i$:
\begin{equation}
    \label{eq_supplydemand}
    p_i + \sum_{j=1}^N a_{ji}f_{ji} = s_i + c_i + \sum_{j=1}^N a_{ij}f_{ij}.
\end{equation}
Here, the sums on the left- and right-hand sides of the equation correspond to the inflow and outflow of cocaine in country $i$, respectively. Data on the nodal amounts of cocaine produced ($p_i$), seized ($s_i$) and consumed ($c_i$) are collected and derived using the approach presented in Section \ref{sec_appendix_marketparametercalculation}. In addition to Equation (\ref{eq_supplydemand}), an additional capacity constraint is imposed, stating that the flow on an individual link in the network should not be larger than the total production amount $P$ in the corresponding year. Lastly, flow can only exist on links present in the transportation network, for which $w_{ij}$ is defined. These constraints reflect the minimal constraints that need to be considered.

\subsection{Trafficking route diversification}
In reality, criminals tend to traffic cocaine over multiple trajectories to spread risk \cite{Magliocca2019ModelingSystem, Europol2025SOCTA}.
Hence, we further generalize our model by adding uniform noise to each link's interception risk. Concretely, given a set of strength parameters $(\alpha_1$,$\alpha_2$,$\alpha_3)$ and noise fraction $\nu\in[0,1]$, the risk of interception on each link will be generalized from $w_{ij}$ to $w_{ij}^*$, a uniformly distributed random variable within the range $[(1-\nu)w_{ij}, (1+\nu)w_{ij}]$. For each realization of the generalized interception risk $\{w_{ij}^*\}$, the corresponding trafficking flow matrix can be derived using Equation (\ref{eq_flowoptimizationproblem}) and the associated constraints. For each link ($i$,$j$), the final trafficking flow $\bar{f}_{ij}$ is estimated as the average flow value across multiple realizations of interception risk.
\\
\\
For each noise fraction, we also propose a baseline model with equal interception risk on all links, i.e., when $\alpha_l=0$ for all $l\in[1,m]$. In the baseline with $\nu=0$, interception risk for every link is equal to one. 
In this case, the purpose of criminals is to minimize the number of network links used to traffic cocaine, analogous to \cite{Leibbrandt2023_DrugNetwork}. 
\\
\\
Given the strength parameter set ($\alpha_1, \alpha_2, \alpha_3$), the interception risk matrix $\{w_{ij}\}$ can be derived for each year using the seven link-level features of that year and Equations (\ref{eq_PC}) and (\ref{eq_w}). Furthermore, interception risk will be generalized as $\{w^*_{ij}\}$, controlled by the noise parameter $\nu$, accounting for uncertainty of interception risk and trafficking diversification. The final trafficking flows $\{\bar{f}_{ij}\}$ of that year can then be identified as the average of those minimizing the total interception cost.

\section{Results}
\label{sec_results}
The calibration of model parameters ($\alpha_1, \alpha_2, \alpha_3$) and $\nu$ will be discussed in Section \ref{sec_performanceanalysis}, while the resulting flow outcomes will be discussed and interpreted in Section \ref{sec_modelinterpretation}. Afterwards, expected flow shifts resulting from changes in law enforcement strategies and perceived interception risk will be analyzed in Section \ref{sec_flowdisplacment}.

\subsection{Performance analysis}
\label{sec_performanceanalysis}
\begin{figure}[h]
\centering
\includegraphics[width=\textwidth]{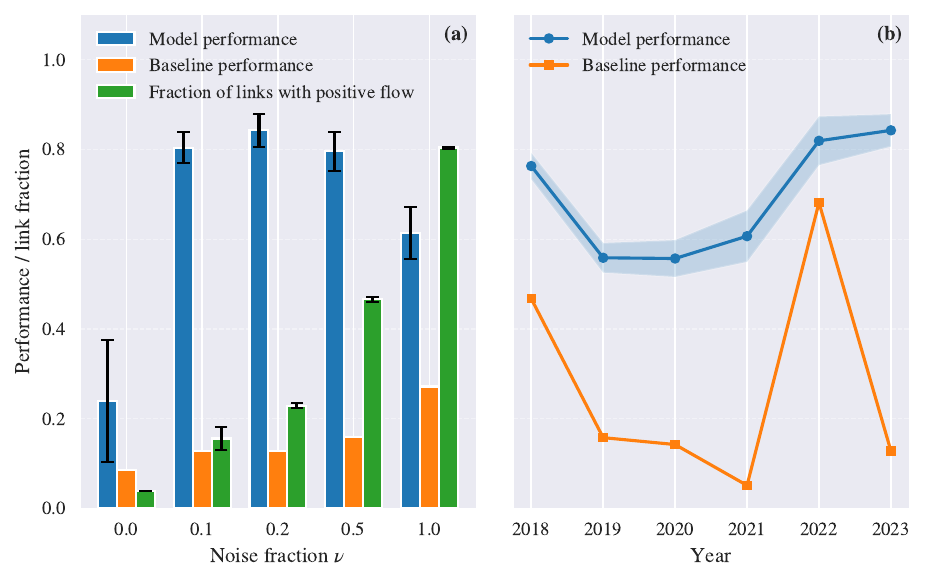}
\caption{\textbf{(a)} Performance of our model (blue) and the baseline model (orange), respectively, and the fraction of links with positive flow in our model (green), for different values of the noise parameter $\nu$ in 2023. \textbf{(b)} Model (blue) and baseline (orange) performance over time for $\nu=0.2$. Error bars and shaded regions denote the standard deviation of the mean values over the 100 best-performing strength parameter sets.}
\label{fig_Performance_vs_Noise_2023}
\end{figure}

The performance of our model with a given a strength parameter set $(\alpha_1, \alpha_2, \alpha_3)$ and a given noise parameter $\nu$ can be evaluated by comparing the resulting flow matrix $\{\bar{f}_{ij}\}$ with known trafficking activity on the transportation network. For this, we have constructed a database of known yearly link-level seizures, e.g., the yearly seizure amount on on each link ($i$,$j$), based on yearly aggregates mentioned in open source reports \cite{EUDA2025EuropeanDrugReport, UNODC2025WorldDrugReport, EUDACocaineAnalysis2022, UNODC2023Cocaine, NationalPoliceNetherlands2022, UNODCEuropol2021CocaineInsights, DEA2020NDTA, WCO2024IllicitTrade, Zeehavenpolitie2025Havenscan1, PJUNCTE2024TCD}. These seizure amounts are converted to pure quantities by multiplying them by the wholesale purity of country $j$.
By using these pure seizure amounts $s_{ij}$ on links ($i$,$j$), the performance of our model with a given strength parameter set and noise parameter can be evaluated as the fraction of correctly identified flow, calculated as
\begin{equation}
    \label{eq_performance}
    \frac{\sum_{(i,j)\in E}\min(s_{ij}, \bar{f}_{ij})}{\sum_{(i,j)\in E}s_{ij}}.
\end{equation}
Figure \ref{fig_Performance_vs_Noise_2023}(a) displays the performance of both our model (averaged over the 100 best performing strength parameter sets) and the baseline in 2023 across different noise levels $\nu$. These best performing strength parameter sets were obtained through a grid search among different strength parameter combinations, where $\alpha_l\in[0.0, 0.2, 0.4,\dots,4.0]$ and $l\in [1,3]$. Besides that, the number of iterations (realizations of the generalized interception risk of links) used to determine the final flow matrix for each strength parameter combination was set to $100$, as model performance was shown to start converging from this number of iterations.
\\
In general, Figure \ref{fig_Performance_vs_Noise_2023}(a) shows that the proposed model significantly outperforms the baselines. The worst performance is obtained by the baseline model with $\nu=0$, where the number of trafficking links is minimized and the interception risks of links are ignored.
The performance of our model first increases and then decreases with increasing $\nu$. This suggests that a moderate noise fraction enables the model to capture criminals' uncertainty about interception risk and the diversity of trafficking routes. In contrast, a high noise level obscures the feature-based interception risk $w_{ij}$ on link $(i,j)$, resulting in poorer model performance. 
\\
\\
Figure \ref{fig_Performance_vs_Noise_2023}(a) also shows that the fraction of network links carrying a positive flow in our model increases with $\nu$. This confirms that introducing noise into the interception risk promotes risk spreading, resulting in greater diversification of trafficking routes.
Among all models, best performance is achieved for $\nu=0.2$, which is the noise fraction that will be considered for the remainder of this work.
Figure \ref{fig_Performance_vs_Noise_2023}(b) displays the performance of both our model and the baseline over time, further showing the outperformance of the proposed model.

\subsection{Model interpretation}
\label{sec_modelinterpretation}
\begin{figure}[h]
\centering
\includegraphics[width=\textwidth]{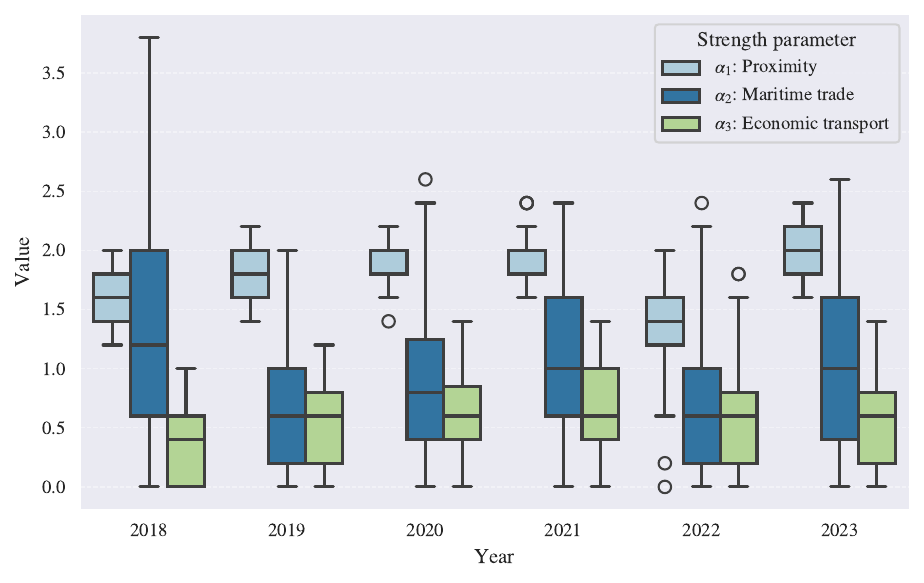}
\caption{Temporal evolution of the distribution of individual strength parameter values among the top 100 best performing parameter sets.}
\label{fig_strength_parameter_comparison_over_time}
\end{figure}

\begin{figure}[h]
\centering
\includegraphics[width=\textwidth]{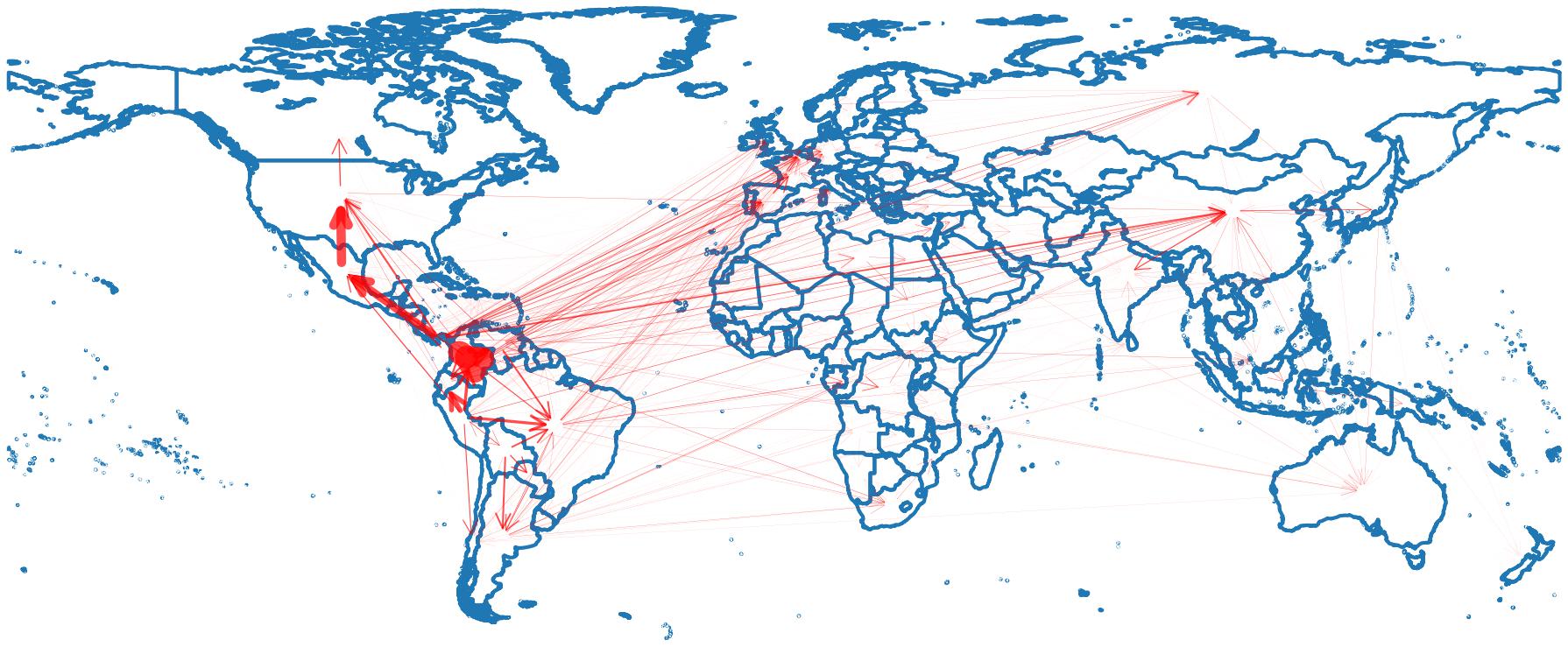}
\caption{Identified flow on the transportation network in 2023, which is the average over the flow matrices corresponding to the 100 best performing strength parameter sets.}
\label{fig_flowconfigexample}
\end{figure}

\begin{figure}[h]
\centering
\includegraphics[width=\textwidth]{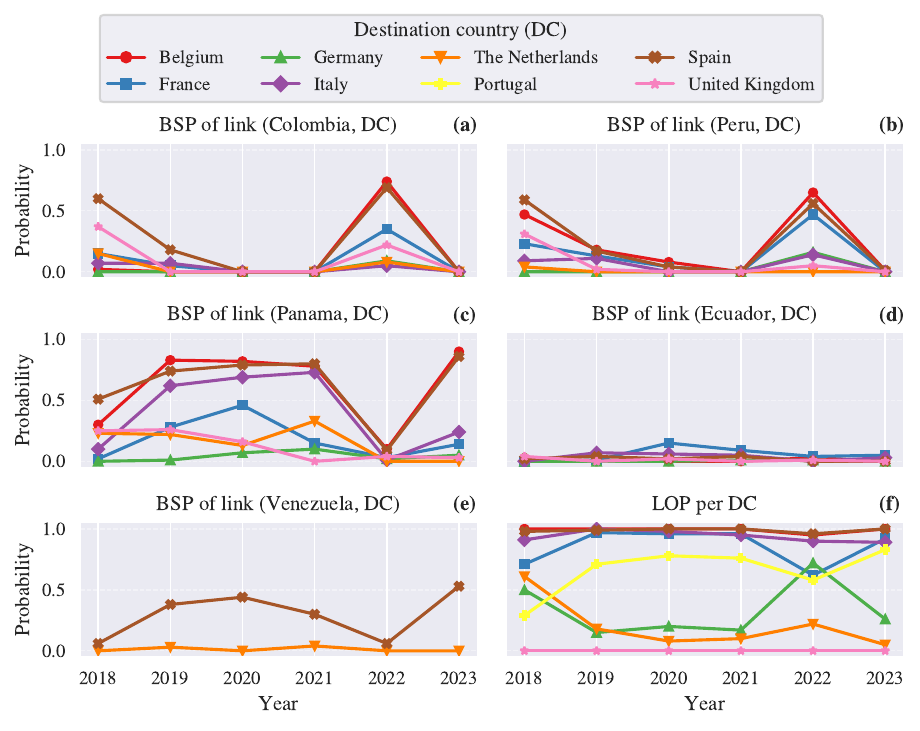}
\caption{\textbf{(a)}-\textbf{(e)} Blind spot probability (BSP) and \textbf{(f)} largest outflow probability (LOP) for links to / countries in Western and Central Europe, respectively, for recent years.}
\label{fig_BS_notseizedflow_EU}
\end{figure}

Figure \ref{fig_strength_parameter_comparison_over_time} summarizes the distribution of the individual strength parameter values appearing in the top 100 best performing sets in each year, suggesting that all three PCs play a role in the perceived risk of interception. Proximity is generally the most prominent driving component. Therefore, links ($i$,$j$) with low proximity scores $y_{1,ij}$ are considered to have a low risk of interception. This finding seems to be consistent with perspective of the local embeddedness of organized crime \cite{Hobbs1998GoingCrime, Leukfeldt2019CriminalEmbeddedness}, which argues that global criminal activities, such as transnational cocaine trafficking, remain rooted in local contexts, infrastructures, and relationships.
\\
\\
The strength parameter related to maritime trade ($\alpha_2$) is the second most prominent among all three parameters. Therefore, links to countries that accommodate a large number of container ships or handle substantial throughput are perceived as carrying low risk of interception. This observation supports that cocaine traffickers exploit port infrastructures and confirms the intersection of legal and illegal trade on container shipping service routes \cite{Sergi2022Pacman}.
Lastly, economic transport ($\alpha_3$) has been shown to only be valued to a lesser extent in the criminal risk analysis. This, together with the observed loadings of the first and third component in Table \ref{tab:pca_compact}, suggests that criminals perceive less risk associated with production proximity and wholesale price differences when trafficking over land connections rather than sea connections.
\\
\\
Based on the 100 best performing sets of strength parameters, we can study how cocaine is likely trafficked around the world. Figure \ref{fig_flowconfigexample} shows the identified flow of cocaine on the transportation network in 2023, as an average over the flow matrices corresponding to the top 100 strength parameter sets. The 100 flow matrices are relatively similar, as any two matrices rank links by flow similarly (the average Spearman rank correlation is equal to $0.70$).
\\
\\
We observe from Figure \ref{fig_flowconfigexample} that the majority of cocaine destined for the United States likely comes from Mexico, followed by Panama, Venezuela, and Colombia, which is in line with observed trafficking patterns derived from the UNODC's IDS database \cite{UNODC2025WorldDrugReport}.
At the same time, Europe seems to be mainly supplied through Panama, Ecuador, and Brazil. 
Although European ports have seized large quantities of cocaine originating from Brazil and Ecuador \cite{EUDACocaineAnalysis2022, EUDA2025EuropeanDrugReport}, cocaine trafficking flows from Panama to Europe, as predicted by our model, remain mostly undetected. Hence, these trafficking routes from Panama to Europe can be identified as `blind spots' in the cocaine supply chain.
\\
\\
To map these blind spots, we study the blind spot probability (BSP) for each link, calculated as the probability for the link to belong to the top five links with the highest undetected flow (i.e., $\bar{f}_{ij}-s_{ij}$) among all 100 flow configurations identified by our model.
Figures \ref{fig_BS_notseizedflow_EU}(a)-(e) show the links incident to Western and Central Europe with the highest BSP as an example.
Here, we again observe the vulnerability of Panamanian shipping lines to be misused for trafficking activities.
An exception is the year 2022, where most undetected flow seems to come directly from the production region.
\\
\\
The majority of the undetected flow appears to be entering Western and Central Europe through Belgium and Spain. A certain part of the flow to these countries will be used to satisfy their internal demand, while the remaining part will be used to be further distributed in Europe. Therefore, to study the transit function of a country, the country's outflow can be analyzed. 
\\
\\
To this end, Figure \ref{fig_BS_notseizedflow_EU}(f) shows the countries in Western and Central Europe with the largest outflow probability (LOP), defined as the probability of appearing in the top five countries with the highest outflow. We observe that Belgium, Spain, and Italy can be considered as the main cocaine gateways to Europe, followed by France and Portugal. Once cocaine has entered through these countries, it is easily distributed to the rest of Europe over land. 
Although single links to Portugal did not appear in Figures \ref{fig_BS_notseizedflow_EU}(a)-(e), the country still appears to be attractive as a transit hub, as many links with smaller flows are used. 
On the other hand, the transit potential of France is mostly due to an internal flow in Europe, where Spain is distributing cocaine to Europe through France. Lastly, we observe that the use of Germany and the Netherlands as cocaine gateways to Europe is declining, with the exception of 2022, and that the LOP of the United Kingdom is zero for all years.
\\
\\
The probabilities in Figure \ref{fig_BS_notseizedflow_EU}(f) can be compared to the country-level transit indications in the UNODC's WDRs \cite{UNODC2025WorldDrugReport}, which are based on seizure information alone. In these reports, the transit roles of Italy and Portugal seem to be underestimated, while the contribution of the Netherlands as a transit hub appears to be overestimated. This suggests that cocaine transports related to the Netherlands are often intercepted, while seizures related to Italy and Portugal possibly represent a small part of the total flow only.

\subsection{Flow displacement effects}
\label{sec_flowdisplacment}
We illustrate further how the proposed model can be used to understand and anticipate different displacements of cocaine trafficking flows. 
First, we can study how flow trajectories are expected to change once cocaine traffickers experience additional law enforcement pressure on certain hubs or routes, a phenomenon also known as the waterbed effect. 
Secondly, we can analyze the displacement of cocaine trafficking flows resulting from a general change in interception risk perceived by criminals, as a direct consequence of how each PC is valued in the link level interception risk analysis.
The prediction of both types of flow displacements could provide new insights and opportunities for law enforcement agencies to combat the illicit cocaine trade.

\subsubsection{Waterbed effect} 
\begin{figure}[h]
\centering
\includegraphics[width=\textwidth]{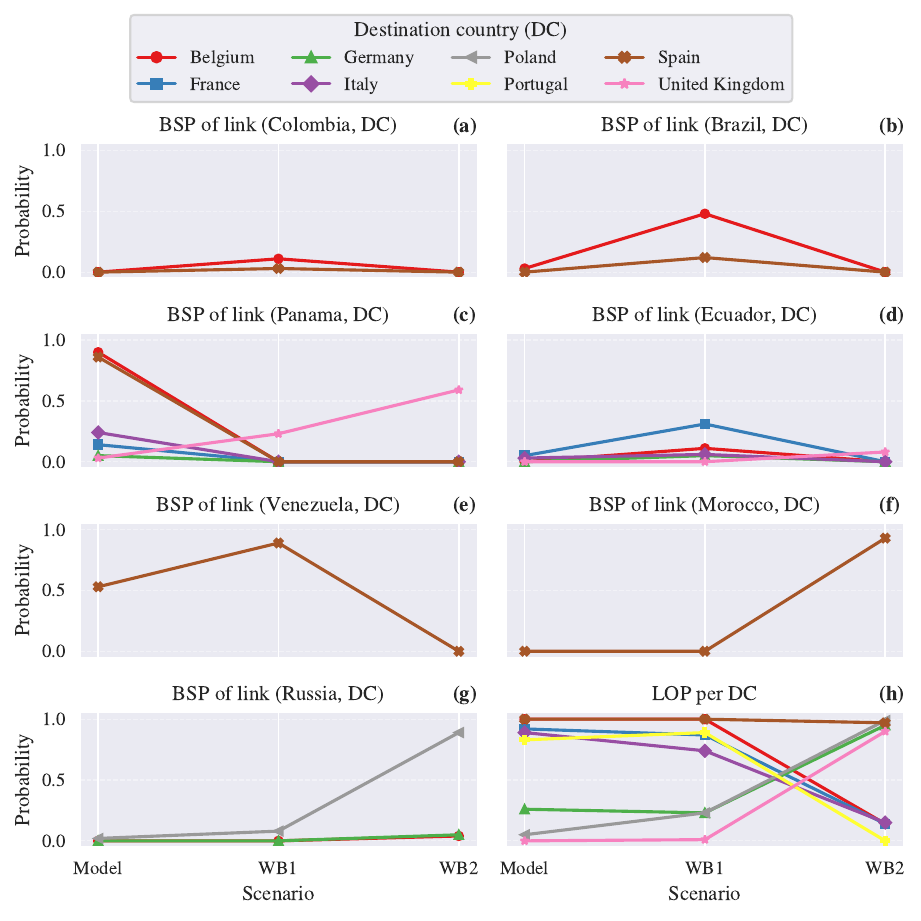}
\caption{\textbf{(a)}-\textbf{(g)} Blind spot probability (BSP) / \textbf{(h)} largest outflow probability (LOP) for links to / countries in Western and Central Europe, respectively, for the model outcome in 2023 and the strategies with additional inspections in Western Europe on links from Panama (WB1) and all countries in the Caribbean and South and Central America (WB2).}
\label{fig_wb}
\end{figure}

The emergence of Panama as a hub for European cocaine, as seen in Figure \ref{fig_BS_notseizedflow_EU}(c), may motivate additional law enforcement inspections on these routes. In turn, this would lead to an increase in interception risk on these links, which might cause these traffickers to seek alternative trajectories carrying less risk of interception. 
We will study the effect of this particular law enforcement strategy, i.e., increasing the inspection probability of ships originating from Panama in Europe. Concretely, this strategy can be mapped into our model by increasing Panama's production proximity risk indicator as $r_i=2$ (in Equation (\ref{eq:customsoriginsamplingrisk})) for links towards Europe's main cocaine entry points (which are Belgium, France, Germany, Italy, the Netherlands, Portugal and Spain, according to Figure \ref{fig_BS_notseizedflow_EU}(f)). Consequently, this leads to a higher law enforcement inspection risk $r_{ij}$ and, in turn, interception risk $w_{ij}$ on the links from Panama to these European countries.
\\
\\
To study the change in blind spot probability (BSP) and largest outflow probability (LOP) for links to / countries in Western and Central Europe, respectively, the cocaine trafficking flows of the top 100 strength parameter sets will be recomputed using the model with increased interception risk on the links from Panama to the Western European coast.
The resulting BSP and LOP for this scenario (denoted as inspection strategy WB1), together with the original model outcome, are shown in Figure \ref{fig_wb} for the year 2023. 
\newpage
\noindent We observe that the BSP for the links with additional law enforcement pressure has decreased to zero percent. While this means that these links are reduced in usage for trafficking, cocaine can still be trafficked undetected on these links in smaller amounts.
As a direct result of the additional enforcement pressure on Panamanian lines, we observe new blind spots appearing, such as the links from Brazil to Belgium and from Ecuador to France. In addition, the route from Venezuela to Spain is likely to be even more exploited by trafficking organizations.
\\
\\
From the strategy of conducting extra inspections on Panamanian shipping lines, we observe that cocaine traffickers are likely to exploit other countries in the Americas as transit hubs for European cocaine.
Therefore, as an additional enforcement strategy, inspections on the Western European coast can be increased on all shipping lines originating from the Americas, in a similar fashion as increasing law enforcement pressure on Panamanian lines alone. The BSP and LOP resulting from this second waterbed scenario are also shown in Figure \ref{fig_wb}, denoted as strategy WB2.
\\
\\
We observe in Figures \ref{fig_wb}(a)-(g) that cocaine is now more likely to enter Europe undetected through links from Morocco to Spain, Russia to Poland, and Panama to the United Kingdom. Moreover, Figure \ref{fig_wb}(h) shows that, with the exception of Spain and Germany, all countries on the Western European coast have disappeared as main cocaine entry points, being replaced by Poland and the United Kingdom. The high LOB of Germany can be explained from the fact that Poland is supplying European cocaine through Germany.
\\
\\
Although only two scenarios of the waterbed effect are presented in this work, many different strategies might be explored by law enforcement agencies themselves.
In the end, the insights gathered from the flow changes as a result of these strategies might prove useful in the allocation of resources in the fight against cocaine trafficking.
\\
\\
Lastly, waterbed effects are not limited to increased law enforcement inspections alone. For example, if the Netherlands becomes more vulnerable to corruption in the future, the Netherlands' CoC value can be correspondingly reduced, thereby decreasing interception risk on all links to the Netherlands. According to the LOP, resulting from the model corresponding to this waterbed scenario, the Netherlands has the potential to become a main cocaine gateway to Europe alongside Belgium and Spain, the main trafficking corridors being from Panama and Venezuela to these countries. 
This finding seems to support the claim that corrupting port workers is essential for smuggling drugs through the port of Rotterdam, the largest container port in the Netherlands \cite{Roks2021GettingRotterdam}.
Ultimately, similar waterbed scenarios related to changes in container ship movements and/or container port throughput (CPT) can be analyzed.

\subsubsection{Shifts in interception risk perception}
\begin{figure}[h]
\centering
\includegraphics[width=\textwidth]{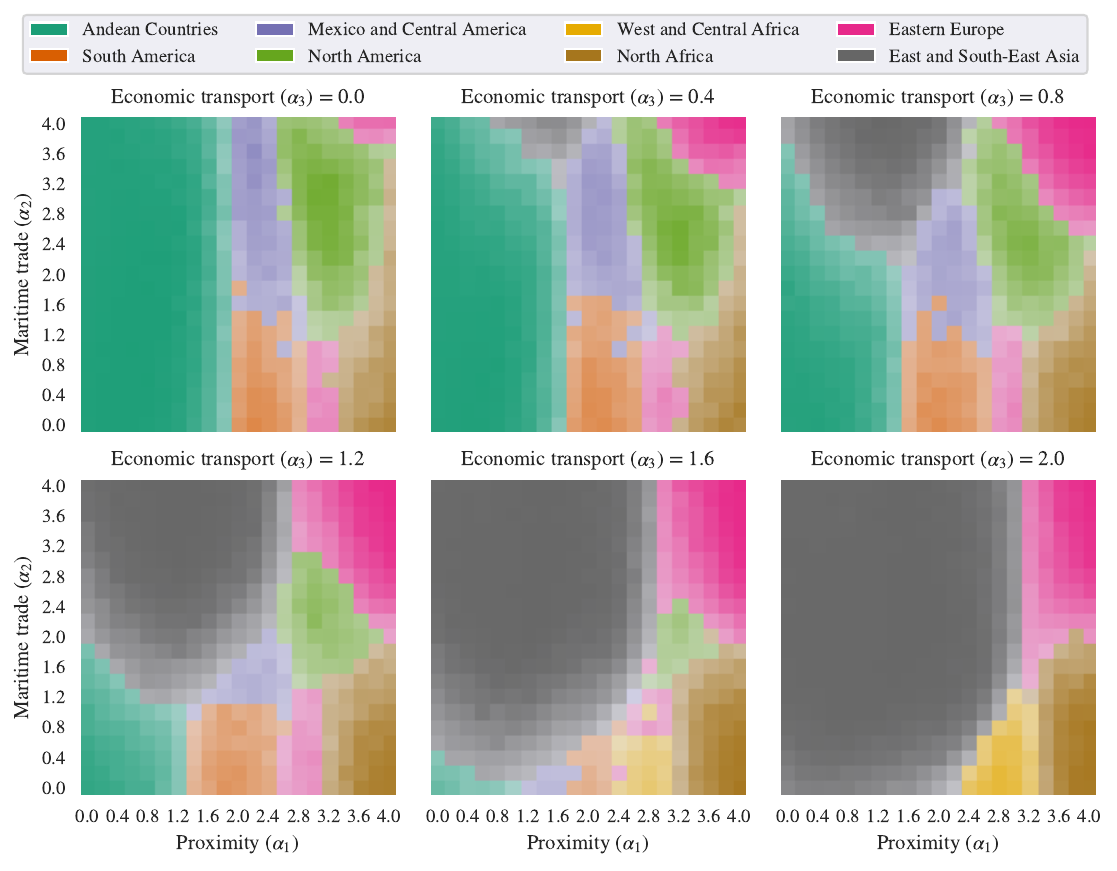}
\caption{Main regions exploited for cocaine trafficking to Western and Central Europe in 2023 for different combinations of strength parameters.}
\label{fig_HM_WCEU23}
\end{figure}

\begin{figure}[h]
\centering
\includegraphics[width=\textwidth]{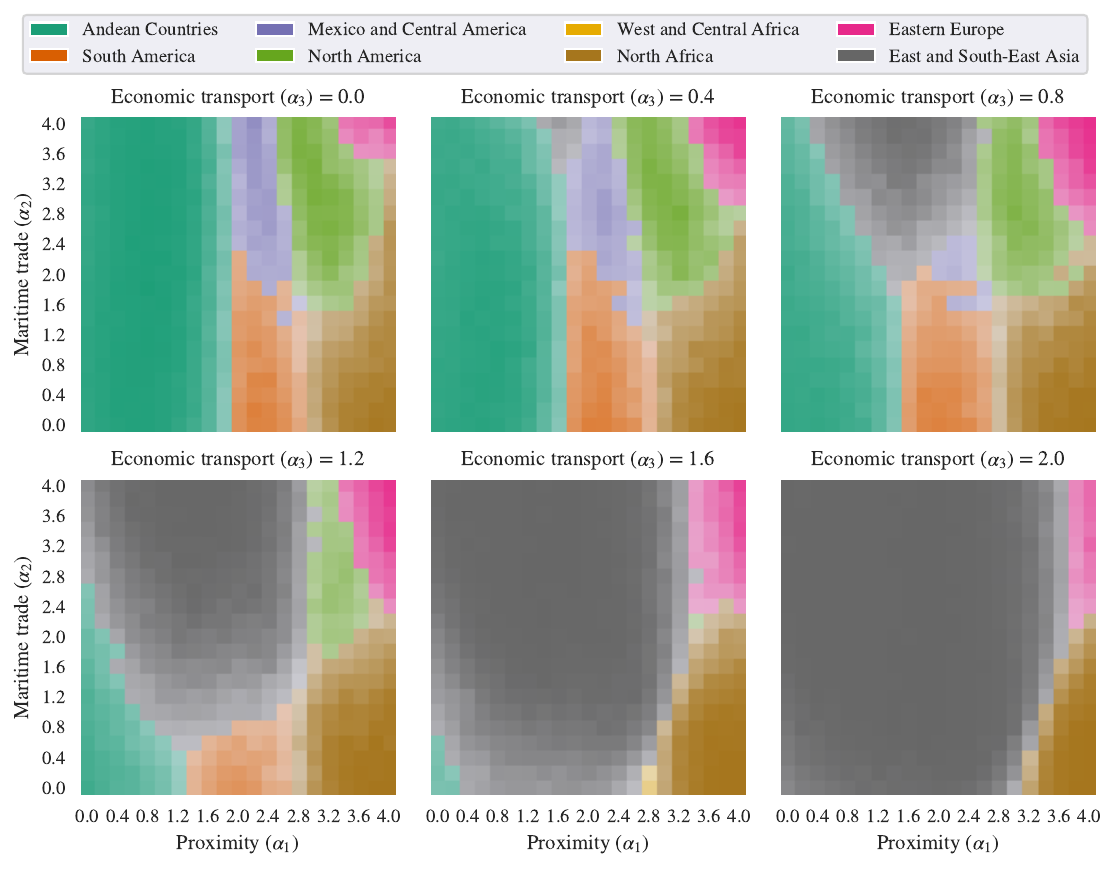}
\caption{Main regions exploited for cocaine trafficking to Western and Central Europe in 2015 for different combinations of strength parameters.}
\label{fig_HM_WCEU15}
\end{figure}

\begin{figure}[h]
\centering
\includegraphics[width=\textwidth]{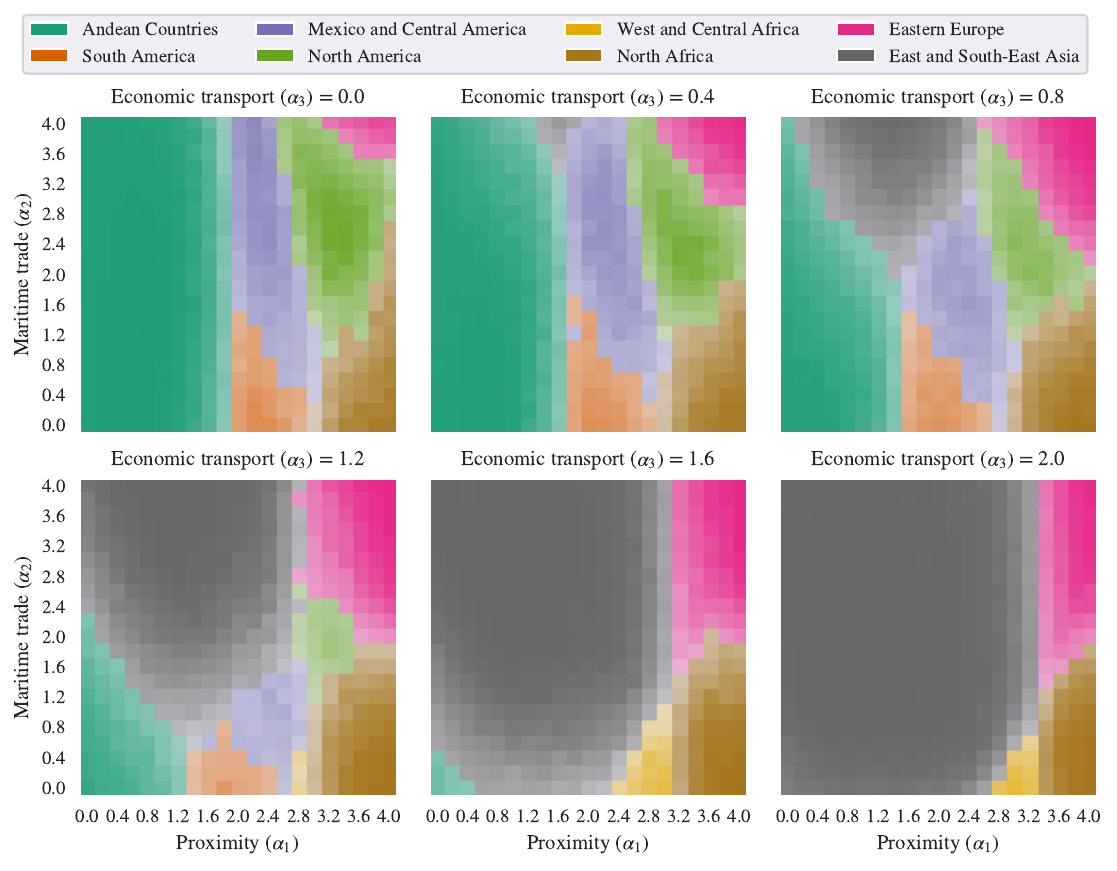}
\caption{Main regions exploited for cocaine trafficking to Western and Central Europe in 2019 for different combinations of strength parameters.}
\label{fig_HM_WCEU19}
\end{figure}

Flow displacements might also occur once cocaine traffickers alter their perception of interception risk, as reflected by changes in the strength parameters. Flow displacements can be observed via the updated flow matrix resulting from the changed parameter values. Without loss of generality, we examine the influence of the strength parameter sets on cocaine routing patterns.
To this end, Figure \ref{fig_HM_WCEU23} shows the regions used for trafficking cocaine to Western and Central Europe directly in 2023, for different sets of strength parameters. For each set, the `main' region is shown (e.g., the region with the largest flow to Western and Central Europe), while the opacity reflects the dominance of the main region (e.g., the percentage of flow coming from the main region).
\\
\\
Based on the top 100 strength parameter sets, South America and Mexico and Central America are shown to be the main transit regions used for the trafficking of cocaine to Western and Central Europe, similarly to the results presented in Section \ref{sec_modelinterpretation}. From Figure \ref{fig_HM_WCEU23} we observe that Central America is preferred as a transit hub for larger $\alpha_2$ values. When proximity is slightly more valued as a driver of interception risk, North America and Eastern Europe have the potential to become a hub for cocaine trafficking to Western and Central Europe. On the other hand, if proximity is valued to a lesser extent, cocaine is more likely to come from the Andean region or East and South-East Asia, the latter being more favored for larger values of $\alpha_2$ and $\alpha_3$. Lastly, when economic transport is more valued than maritime trade as driver of interception risk, West and Central Africa becomes a potential hub for trafficking cocaine to Europe. The importance of this region appears to increase with time, as shown in Figures \ref{fig_HM_WCEU23}, \ref{fig_HM_WCEU15} and \ref{fig_HM_WCEU19}, an observation that seems to be consistent with previous studies \cite{INTERPOL2023CocaineAfrica, VanDenEeden2026GeorganiseerdeCriminaliteit}.

\section{Conclusion}
\label{sec_conclusion}
In this work, we proposed a novel framework to analyze global cocaine trafficking flows using a network-based model. This model assumed that cocaine is routed from production countries to consumption countries on a transportation network of land and sea connections, by minimizing the total risk of interception by law enforcement, independent of seizure information. Instead, the interception risk of each link, perceived by cocaine traffickers rationally, depended on diverse features such as container ship movements, whether the link is a land connection or not, and the vulnerability of the link's target country to corruption.
Using Principal Component Analysis, all features were mapped into three key drivers of interception risk. The importance of each driving component in the calculation of interception risk was characterized by a corresponding strength parameter. For a given combination of the three strength parameters, the model determined cocaine trafficking flows, defining the exact volume transported along each link. 
The model was validated using link-level seizure data, adhering to the principle that the modeled trafficking flow on any given link is supposed to exceed the volume seized along that link. Across the 100 best performing strength parameter sets, our model outperformed baseline models such as those that assume a uniform interception risk across all links.
\\
\\
We demonstrated how this model could be used to 1) identify blind spots, i.e., links where the model's estimated volume of trafficked cocaine significantly exceeds the amount intercepted, 2) anticipate the waterbed effect, i.e., how enhanced law enforcement at selected routes results in the displacement of cocaine trafficking flows, and 3) anticipate the displacement of cocaine trafficking flows due to shifts in interception risk perceived by criminals, i.e., the strength parameters. These insights could facilitate the development of effective law enforcement strategies. 
\\
\\
Our preliminary model that identifies actual cocaine trafficking flows is confined to certain assumptions and could thus be further improved. 
Firstly, the cost of cocaine trafficking is not necessarily attributable to interception risk alone. Operational costs related to, for example, the bribing of port workers, criminal interference in ports, and incarceration risk, could be included in the flow optimization objective. Secondly, it is assumed that all cocaine produced in a given year is consumed within that same year. In practice, however, a portion of annual production may be stockpiled for consumption in subsequent years. Thirdly, more transport modalities could be considered in both network construction and interception risk determination. Smuggling in the Caribbean, for example, is also done using planes, go-fast boats and/or fishing boats to avoid law enforcement controls \cite{DEA2020NDTA, UNODC2023Cocaine}.
The quality of both the underlying UNODC data as well as the link-level seizure data could be further improved, as it is accompanied by several limitations, such as reporting consistency, bias, and missing data \cite{Aziani2018IllicitEstimation, Bichler2023DrugStructure}. To this end, nodal reporting rates and different imputation methods could be considered. 
Lastly, the framework presented in this work could also be applied and further adapted to understand other types of trafficking, such as trafficking in other drugs, wildlife, and human beings.

\section{Acknowledgements}
We thank for the support of Netherlands Organisation for Scientific Research NWO (TOP Grant no. 612.001.802, project FORT-PORT no. KICH1.VE03.21.008).

\section{Data availability statement}
The data that support the findings of this study are openly available and correctly referred to within the text.

\section{Disclosure statement}
No competing interest was reported by the authors.

\printbibliography
\section{Appendix}
\label{sec_appendix}
\subsection{Consumption calculation}
\label{sec_appendix_marketparametercalculation}
In order to reconstruct cocaine trafficking flows, the annual size of each country's cocaine market can be characterized by its levels of production, seizures and consumption, analogous to the method of Giommoni et al. \cite{Giommoni2022_InterdictingInterventions}. We estimate these quantities based on the variables shown in Table \ref{tab:variablesMarket}, for which we obtained data for the years $2010$-$2023$ from the UNODC's WDRs \cite{UNODC2025WorldDrugReport}. 
\begin{table}[h]
\centering
\begin{threeparttable}
\caption{Country-level data used in the calculation of the cocaine market parameters.}
\label{tab:variablesMarket}
\begin{tabularx}{\linewidth}{l X}
\hline
\textbf{Variable} & \textbf{Description} \\
\hline
Cultivation area$^{*}$ & Total area under coca bush cultivation for Colombia, Peru and Bolivia. \\
Cocaine manufacture$^{*}$ & Total amount of cocaine produced globally. \\
Seizures (reported)$^{*}$ & Reported amount of cocaine seized (not adjusted for purity). \\
Wholesale purity$^{*}$ & Purity of cocaine on the wholesale level. \\
Prevalence$^{*}$ & Percentage of a country's working population that has used cocaine last year (regional estimates also available). \\
Population$^{\dagger}$ & Total and working population figures. \\
\hline
\end{tabularx}
\begin{tablenotes}
\footnotesize
\item[*] Source: UNODC \cite{UNODC2025WorldDrugReport}.
\item[$\dagger$] Source: World Bank (Worldwide Governance Indicators) \cite{WorldBank2025WGI}.
\end{tablenotes}
\end{threeparttable}
\end{table}
\\
For each year, the production $p_i$ of country $i$ is firstly estimated from the total amount $P$ of pure cocaine manufactured globally. Taking the assumption that the amount of cocaine produced in a country is proportional to its area under coca bush cultivation, $p_i$ can be calculated as the product of $P$ and the country's share of total coca bush cultivation.
The WDRs only report estimates of coca bush cultivation in Colombia, Peru, and Bolivia, which are also the three main cocaine producing countries \cite{UNODC2025WorldDrugReport}. Hence, cocaine is assumed to be produced only in these three countries, which will be referred to as production countries.
\\
\\
Secondly, the amount of cocaine seized in each country is (partially) recorded yearly \cite{EUDA2024StatisticalBulletin, UNODC2025WorldDrugReport}. However, the purity of these seizures is unknown, making it impossible to directly compare them with production estimates. As cocaine is seized mainly in bulk at the wholesale level of the supply chain, the pure amount of cocaine per country can be calculated as the product of each country's wholesale purity and reported seizure volume.
Not all countries report their annual drug statistics to the UNODC, meaning that missing values have to be imputed. In the case of cocaine seizures, missing values are assumed to be zero if no seizure is reported. Missing purity values are imputed through linear interpolation and extrapolation of each country's purity time series. If the purity of a country is not reported for any year, it is estimated as the weighted (by the working population) average of the purities reported by countries in that year within the same region. Regional classification is defined by the UNODC \cite{UNODC2025WorldDrugReport}. The working population (aged 15-64) per country is taken from World Bank's Worldwide Governance Indicators \cite{WorldBank2025WGI}, which is augmented by gathering missing values from local governments. 
\\
\\
Thirdly, the consumption $c_i$ of pure cocaine in country $i$ is estimated for each year. While accurate data on pure cocaine consumption can be obtained from wastewater measurements in cities, the number of consistently reporting cities over time remains limited, despite a clear upward trend in the number of reporting cities \cite{EUDA2025WastewaterAnalysis}. In addition, estimating national consumption from city-level measurements alone poses significant methodological challenges, such as difficulties in deriving national estimates from city-level measurements \cite{Jensen2022KeyStudy, HUIZER2021117789}. For these reasons, $c_i$ is determined using an approach that combines data from both the supply and demand side.
\\
\\
The worldwide consumption of pure cocaine in a given year is estimated as $P - \sum_j^Ns_j$, where $s_j$ is the amount of pure cocaine seized in country $j$. Hence, it is assumed that all annual produced cocaine will be either seized or consumed, and that the amount $c_i$ of pure cocaine consumed in the country $i$ in a year is proportional to the number of cocaine users $u_i$ in this country in that year. Hence,
\begin{equation}
    c_i = (P - \sum_j^Ns_j)\cdot\frac{u_i}{\sum_j^N u_j},
\end{equation}
where the number of users $u_i$ in country $i$ is estimated as the product of its working population and prevalence, which is the percentage of the working population that uses cocaine in a given year \cite{EUDA2025EuropeanDrugReport}. As prevalence is not always reported by UN members, missing values are imputed according to the same method used to impute the missing wholesale purity data, the only difference being that the regional averages already given in the WDRs are used.
The pure amounts of cocaine produced ($p_i$), seized ($s_i$) and consumed ($c_i$) in each year can now be used to reconstruct cocaine trafficking flows.

\newpage
\subsection{Trafficking hub heatmaps}

\end{document}